\documentclass[%
 reprint,
showpacs,
 amsmath,amssymb,
 aps,
]{revtex4-1}

\usepackage{graphicx}
\usepackage{dcolumn}
\usepackage{bm}

\usepackage{graphics}
\usepackage{bm}
\usepackage{amsmath}
\usepackage{epstopdf}
\usepackage{amsmath}
\usepackage{amsfonts}
\usepackage{color, colortbl}
\usepackage[]{algorithm2e}
 \usepackage{tikz}
 \usepackage{mathtools}
 \usepackage[percent]{overpic}
 \usepackage{float}
 \usepackage{hyperref}
\let\MYoriglatexcaption\caption
\renewcommand{\caption}[2][\relax]{\MYoriglatexcaption[#2]{#2}}


\renewcommand{\eqref}[1]{(\ref{#1})}

\newcolumntype{d}[1]{D..{#1}}

\definecolor{Gray}{gray}{0.8}

\begin{document}

\preprint{APS/123-QED}

\title{Promoting cooperation by preventing exploitation: The role of network structure} 

\author{Zoran Utkovski$^{1}$}
\altaffiliation[This work was done while the author was at ]{Macedonian Academy of Sciences and Arts and Faculty of Computer Science, University Goce Delcev, Stip, Republic of Macedonia.} 
\author{Viktor Stojkoski$^{2}$}
\author{Lasko Basnarkov$^{2,3}$}
\author{Ljupco Kocarev$^{2,3}$}

\affiliation{
$^{1}$Fraunhofer Heinrich Hertz Institute, Einsteinufer 37, 10587, Berlin, Germany}
\affiliation{
$^{2}$Macedonian Academy of Sciences and Arts, P.O. Box 428, 1000 Skopje, Republic of Macedonia}%
\affiliation{
$^{3}$Faculty of Computer Science and Engineering, Ss. Cyril and Methodius University, P.O. Box 393, 1000 Skopje, Republic of Macedonia\\}%


\date{\today}

\begin{abstract}

A growing body of empirical evidence indicates that social and cooperative behavior can be affected by cognitive and neurological factors, suggesting the existence of state-based decision-making mechanisms that may have emerged by evolution. Motivated by these observations, we propose a simple mechanism of anonymous network interactions identified as a form of generalized reciprocity -- a concept organized around the premise ``help anyone if helped by someone'', and study its dynamics on random graphs. In the presence of such mechanism, the evolution of cooperation is related to the dynamics of the levels of investments (i.e. probabilities of cooperation) of the individual nodes engaging in interactions. We demonstrate that the propensity for cooperation is determined by a network centrality measure here referred to as \textit{neighborhood importance index} and discuss relevant implications to natural and artificial systems. To address the robustness of the state-based strategies to an invasion of defectors, we additionally provide an analysis which redefines the results for the case when a fraction of the nodes behave as unconditional defectors.
\end{abstract}

\pacs{87.23.Ge, 
87.23.Kg, 
02.50.Ey, 
02.50.Le 
} 
\maketitle

\section{Introduction}

Cooperation has played a fundamental role in many of the major transitions in biological evolution and is essential to the functioning of a large number of biological systems \cite{Axelrod-2006}. Cooperative interactions are required for many levels of network organization ranging from single cells to groups of animals and, ultimately, humans. 

While kin selection \cite{Hamilton-1964} and group (multilevel) selection \cite{Williams-1957, Traulsen-2006} have been able to explain the emergence and stability of cooperation in related individuals, the occurrence of cooperation between unrelated individuals is more intriguing \cite{West-2007, Clutton-2009}. Theoretical models provide evidence that cooperative behavior can nevertheless evolve and persist if it is based on reciprocity \cite{Trivers-1971, Nowak-2006five, Lehmann-2006}. An important role in the emergence of cooperation plays the network structure. Since the early work of Novak and May \cite{Nowak-1993} which demonstrated that a lattice structure enhances cooperation in a Prisoner's Dilemma (PD) game, this issue has attracted a great deal of attention, see for example \cite{ Nowak-2004, Szabo-2004, Lieberman-2005, Pacheco-2006, Rong-2007, Morita-2008, Wang-2014, Skanata-2016}. In particular, the consequences of population structure on the evolution of reciprocal cooperation were studied in \cite{Ohtsuki-2006, Santos-2006, Taylor-2007, Rankin-2009}. It has also been recognized that underlying network structures such as network heterogeneity, scale-freeness etc., crucially determine the outcome of multiple dynamical phenomena \cite{Dorogovtsev-2008}.

To what extent reciprocity can explain the behavior of biological organisms is a subject of active debate (see e.g. \cite{vanDoorn-2012, Taborsky-2016} for a recent review). According to \cite{vanDoorn-2012}, a major concern is that the assumptions of theoretical models differ in important ways from the observed structure of real interactions, as supported by experimental evidence \cite{Hauert-2002, Croft-2008, Connor-2010}. As a result, the mechanisms proposed in some of the theoretical models are unlikely to be realized by evolution in real organisms \cite{Clutton-2009}. Particularly, direct and indirect reciprocity require cognitive abilities to register identity of social partners and their behavior in previous interactions, which has been shown to constrain cooperation in animals \cite{Stevens-2005}, including humans \cite{Milinski-1998}.

Recent empirical studies have shown that cooperation in animals (rats \cite{Rutte-2007}, monkeys \cite{Leimgruber-2014}, dogs \cite{Gfrerer-2017}), as well as humans \cite{Bartlett-2006,Stanca-2009}, can work between non-relative conspecifics by \textit{generalized reciprocity} - a simple mechanism which does not require higher cognitive demands. This mechanism, simply described as ``help anyone if helped by someone'', assumes that an individual who received help in the past is more likely to help any new individual in subsequent interactions. Generalized reciprocity, which can be traced back to ``upstream tit-for-tat'' \cite{Boyd-1989} and ``upstream indirect reciprocity'' \cite{Nowak-2005}), has been recently addressed in detail in \cite{vanDoorn-2012, Taborsky-2016}.

A growing body of recent empirical research indicates that social and cooperative behavior can be affected by cognitive and neurological factors, such as experience and hormone titres \cite{Fox-2009}. In this context, it has been suggested  that the proximate mechanism of generalized reciprocity is based on changes of the individuals’ physiological/neurological state \cite{Rutte-2007,Bartlett-2006, Isen-1987}. 
The first steps in the direction of understanding the evolutionary processes underlying generalized reciprocity have been made in \cite{Boyd-1989,Nowak-2005}. In \cite{Barta-2011} the formation of a decision-making mechanism based on an \textit{internal state} has been investigated by evolutionary simulation. There it has been demonstrated that a mechanism where the individuals base their decision to cooperate based on a state variable updated by the outcome of the last interaction with an anonymous partner, can emerge through small evolutionary steps under a wide range of conditions.

In the presence of supporting empirical evidence for the evolutionary development of a state-based decision-making mechanism, we propose a general framework to address the dynamics of such mechanisms on complex networks. We adopt a simple (stochastic) model for network interactions, where nodes regularly send cooperation requests to randomly chosen neighbors. The selected nodes accept the requests (cooperate) with probability determined by a single variable - an \textit{internal cooperative state} which reflects their current ``well being''. The resulting behavioral mechanism relates the nodes' behavior in the network interactions to their fitness, i.e. accumulated payoff in a game-theoretic jargon, with immediate implications to a large plethora of real-life networks. 

From a game-theoretic perspective, the behavioral mechanism we address may be framed in the \textit{continuous Prisoner dilemma} context \cite{Wahl-1999I} according to which the level of investment (i.e. the probability of cooperation) is adjusted to the accumulated experience. While in our stochastic model interactions happen between two individuals, the fact that the behavioral mechanism is oblivious to the identity of donors and receivers effectively provides a framework where the network nodes engage in a game with their neighborhood. This, in a sense, is conceptually similar to some versions of the N-player iterated Prisoner’s Dilemma \cite{Doebeli-2005,ORiordan-2008}. 

We point out that in our model we do not consider  ``competition'' between strategies in the sense of \cite{Axelrod-1981}. Also, we do not assume evolutionary updates in the sense of e.g. \cite{Nowak-2005}, or imitation of the neighbor's strategy (e.g. imitation dynamics) \cite{Smith-1973}. Instead, we presume that a general form of state-based generalized reciprocity mechanism is in place in the complex network of interest (i.e. has evolved as a result of evolution, as suggested by empirical evidence), effectively resulting in a continuum of investment strategies with levels of investment changing according to the nodes individual states. 

With this in mind, the aim of our approach is to evaluate the implications of this behavioral mechanism on the cooperation in the network and, importantly, to reveal the role of network structure. In particular, we show that, in the presence of the here addressed state-based behavioral mechanism, the levels of individual investments may evolve from very low to significant, eventually saturating to a point which is solely determined by the network topology. To address the robustness of the state-based strategies to an invasion by  defectors, we restate the results for the case when a fraction of the network nodes behave as unconditional defectors. 
	
An important aspect of our model is the fact that the anonymity of the network interactions is not associated with an increased vulnerability to exploitation (as it is generally the case with generalized reciprocity \cite{vanDoorn-2012,Taborsky-2016}. In fact, we are able to show that the behavioral mechanism promotes cooperation by ``driving'' the network towards a steady state beyond which the individual nodes are protected from exploitation by the rest of the network.

The remaining of the paper is structured as follows. In Sec.~\ref{sec:Model} we define the stochastic model for network interactions and describe the proposed behavioral mechanism (update rule). In the same section we present the deterministic counterpart of the stochastic model, which is analytically tractable and represents a valid approximation in the steady state regime.  In Sec.~\ref{sec:Results} we analyze the deterministic model and address the issue of cooperation from the perspective of the network topology. We further derive important properties of the model related to the spread and stability of cooperation in the network. The analytical findings are supported by numerical results. We conclude the paper by discussing the implications of the model and drawing parallels with other theoretical models and real-life networks.

\section{Model Description}
\label{sec:Model}

\subsection{Network model}
\label{sec:Network_Model}
The network is modeled as a random graph $\mathbf{A}$ on a finite set $\mathcal{N}$ of $N$ nodes, with binary edge variables $A_{ij} \in \{0,1\}$ between pairs of nodes $i, j \in \mathcal{N}$ ($ A_{ij}=1, i\neq j$ indicating neighborhood relation). The interactions between the nodes are modeled as follows: in each round $t$, node $i$  sends a cooperation request to a randomly (on uniform) chosen node from its neighborhood, for example $j\in\mathcal{N}_i$; upon selection, node $j$ accepts the request (i.e. cooperates) with probability $\mathrm{p}_j(t)$, which may be considered as its  internal \textit{cooperative state} at time $t$; if node $j$  accepts the request, i.e. cooperates, it pays a cost $c$ for node $i$ to receive a benefit $b$ (we assume these quantities to be the constant over the network). The (random) payoff of node $i$ at round $t$ is
\begin{align}
\mathrm{y}_i(t)&=b \mathrm{x}_{j}(t) -c \mathrm{x}_i(t)\sum_{k\in\mathcal{N}_i} \rho_{k}(t).
\label{eq:payoff}
\end{align}
In (\ref{eq:payoff}): the selected index from the neighborhood of $i$ is a random variable uniformly distributed on the set $\mathcal{N}_i$, $j\sim \mathrm{U}(\mathcal{N}_i)$; $\mathrm{x}_l(t)$, $l=1,\ldots, N$, are Bernoulli random variables, each with parameter $\mathrm{p}_l(t)$ (the cooperative state); $\rho_h$ is a Bernoulli random variable with parameter $1/d_h$, where $d_h$ is the degree of node $h$, $d_h=\sum_l A_{h,l}$; the term $\sum_{h\in\mathcal{N}_i} \rho_{h}(t)$ captures the (random) number of nodes (neighbors of $i$) which send a cooperative request to $i$ during round $t$. We note that the model (\ref{eq:payoff}) may be easily extended to weighted graphs by substituting the uniform distribution with categorical.

\subsection{Behavioral mechanism (update rule)}
For simplicity, we assume a synchronous behavioral update, based on of the accumulated (i.e. total) payoff of the node $i$ by time $t$, $\mathrm{Y}_i(t)=\mathrm{Y}_i(t-1)+\mathrm{y}_i(t)$, with $\mathrm{Y}_i(0)$ being the initial condition and $\mathrm{y}_i(0)=0$.  
The cooperative state of node $i$ at time $t+1$ is defined as 
\begin{align}
\mathrm{p}_i(t+1)=\mathrm{f} \left(\mathrm{Y}_i(t)\right),
\label{eq:update}
\end{align} 
where we assume that the function $\mathrm{f}:\mathbb{R}\rightarrow [0\:\:1]$ is monotonic (nondecreasing). A plausible choice which reflects real-world behavior is the sigmoid (logistic) function 
\begin{align*}
\mathrm{f}(\omega)=\left[1+e^{-\kappa (\omega-\omega_0)}\right]^{-1},
\end{align*}
where the parameters $\kappa$ and $\omega_0$ define the steepness, respectively the midpoint of the function. 

We note that it is straightforward to extend the model to account for an asynchronous behavioral update, where in each step $t$ node $i$ updates its probability of cooperation with probability $u$. In that case, the cooperative state of node $i$ at time instant $t+1$ is defined as 
\begin{align*} 
\mathrm{p}_i(t+1)=\mathrm{p}_i^{1-\lambda}(t) \cdot \mathrm{f}^{\lambda}\left(\mathrm{Y}_i(t)\right).
\end{align*}
The dynamics of the behavioral update, 
is then dictated by the payoff accumulated by each node in the (random) time period between two updates $t_o$ and $t_o+T_i$
\begin{align*}
& \Delta \mathrm{Y}_i(t_o, t_o+T_i)\doteq \mathrm{Y}_i(t_o+T_i)-\mathrm{Y}_i(t_o)\\
&=b\sum_{\tau=t_o+1}^{t_o+T_i} \mathrm{x}_{\mathrm{j}}(\tau) -c \mathrm{x}_i(t_o+1)\sum_{\tau=t_o+1}^{t_o+T_i} R_i(\tau),
\end{align*}
with the comment that the index $j\in\mathcal{N}_i$ is updated in each step $\tau$. 

In the following we will only address the scenario  with synchronous update, with the remark that the conclusions also apply to the asynchronous scenario.

\subsection{Random graphs}
The starting point for studying games on graphs are the models used in evolutionary biology, where the evolution of the population over time can be determined by solving a coupled set of differential equations (the replicator equations, see, e.g. \cite{Hofbauer-1998}). Besides being deterministic (no stochasticity in the decisions), this framework assumes infinite, well-mixed populations. 

To account for stochastic game dynamics and finite populations, evolutionary graph theory provides a mathematical tool for representing population structure: nodes correspond to individuals and
edges indicate interactions \cite{Ohtsuki-2006}. Graphs can describe spatially structured populations
of bacteria, plants or animals, tissue architecture and differentiation
in multi-cellular organisms, or social networks. In this context, the well-mixed population, which is a classical scenario for mathematical studies of evolution, is given by the complete graph.

In this setting, the structure of the underlying random graph dictates the final result of many real world systems, including cooperation. In general, real world networks are characterized with three properties \cite{Newman-2003}: i) high clustering - two nodes have a higher probability to share an edge if they have similar neighborhoods, ii) small-world - short, on average, distance (shortest path length) from one node to another, and iii) scale-freeness - power law degree probability density function (pdf). 

Many models have been developed for generating random graphs that have (some of) these properties. We study the behavior of our model on the four models that are most often implemented i) Random $d$-regular graph, ii) Erdos-Renyi (ER) random graph, iii) Watts-Strogatz (WS) random graph, and iv) Barabasi-Albert random scale-free network. In the following, we describe them.

$\bullet$ Random $d$-regular graph -- the simplest random graph that can be found in the literature. Formally defined as a random graph $ \mathbf{A}\left(N, d \right)$ in which all nodes have the same degree $d$ \cite{Bollobas-2013}. As such, it has a degree pdf (and hence a $z$ pdf) described with the Dirac delta function, whereas clustering and shortest path length generally depend on the parameter $d$. To generate a $d$-regular graph we implement the pairing algorithm described in \cite{Kim-2003}. 

A special type of a regular graph that has been commonly studied in evolutionary biology is the two dimensional (2D) $N \times N$ square lattice \cite{Nowak-1993}. A such lattice is characterized with low (zero) clustering and long average path length. The main difference between a 2D square lattice and other random regular graphs is that the structure of the former is not random. Namely, in it the nodes are distributed at the integer coordinate points of the two dimensional Euclidean space and each node is connected to other nodes that are one unit away from it.

$\bullet$ Erdos-Renyi (ER) random graph -- also known as the $ \mathbf{A}\left( N, \pi \right)$ model \cite{Erdos-1960}. In it, two nodes share an edge with probability $\pi$, independently from the presence of other edges. A random graph constructed through this algorithm is characterized with very low clustering, long shortest path length and Poisson degree pdf.

$\bullet$ Watts-Strogatz (WS) random graph -- a model for generating random graphs introduced in \cite{Watts-1998} and defined as $ \mathbf{A}\left( N, d, \beta \right)$, where $d$ is the average degree and $\beta \in \left[ 0, 1\right]$ is the probability that an edge will be ``rewired''. In short, the construction of a WS random graph is as follows. First, a $d$-regular ring lattice is constructed by putting the nodes on integer values of a circle with circumference $N+1$ and connecting them to their $d$ nearest neighbors. Then, each generated edge $\left( i, j \right)$ (with $i < j$) is rewired to $\left( i, k \right)$, where $k \neq i$ is a uniformly chosen node, with probability $\beta$. We point out that $d$-regular ring lattices and ER random graphs emerge as special types of the WS graph when $\beta = 0$ and $\beta = 1$, respectively. When $0 < \beta < 1$ the existence of a local ring lattice structure produces high clustering, whereas the randomly reallocated edges lead to short path lengths, i.e. the small-world property.

$\bullet$ Barabasi-Albert (BA) random scale-free network -- a model based on the preferential attachment mechanism for generating random graphs \cite{Barabasi-1999}. The construction of a BA network, written as $ \mathbf{A}\left(N, m\right)$, is represented as a dynamical process. Concretely, in the beginning a fully connected network of $m_0$ nodes ($\mathcal{N}_{i}=\mathcal{N} \setminus i$ for all $i$) is created. Then, at each time step a new node $i$ is born that makes connections to $m$ other nodes that are present in the network. The node connects to a particular node $j$ with probability proportional to its current degree. Besides having the same properties of high clustering and small-world as the WS graph, the BA graph has a scale-free degree pdf. Therefore, the BA model has been extensively applied for studying real world systems, ranging from social to biological networks and beyond \cite{Barabasi-1999, Barabasi-2004}.

\begin{figure*}[t!]
\includegraphics[width=17cm]{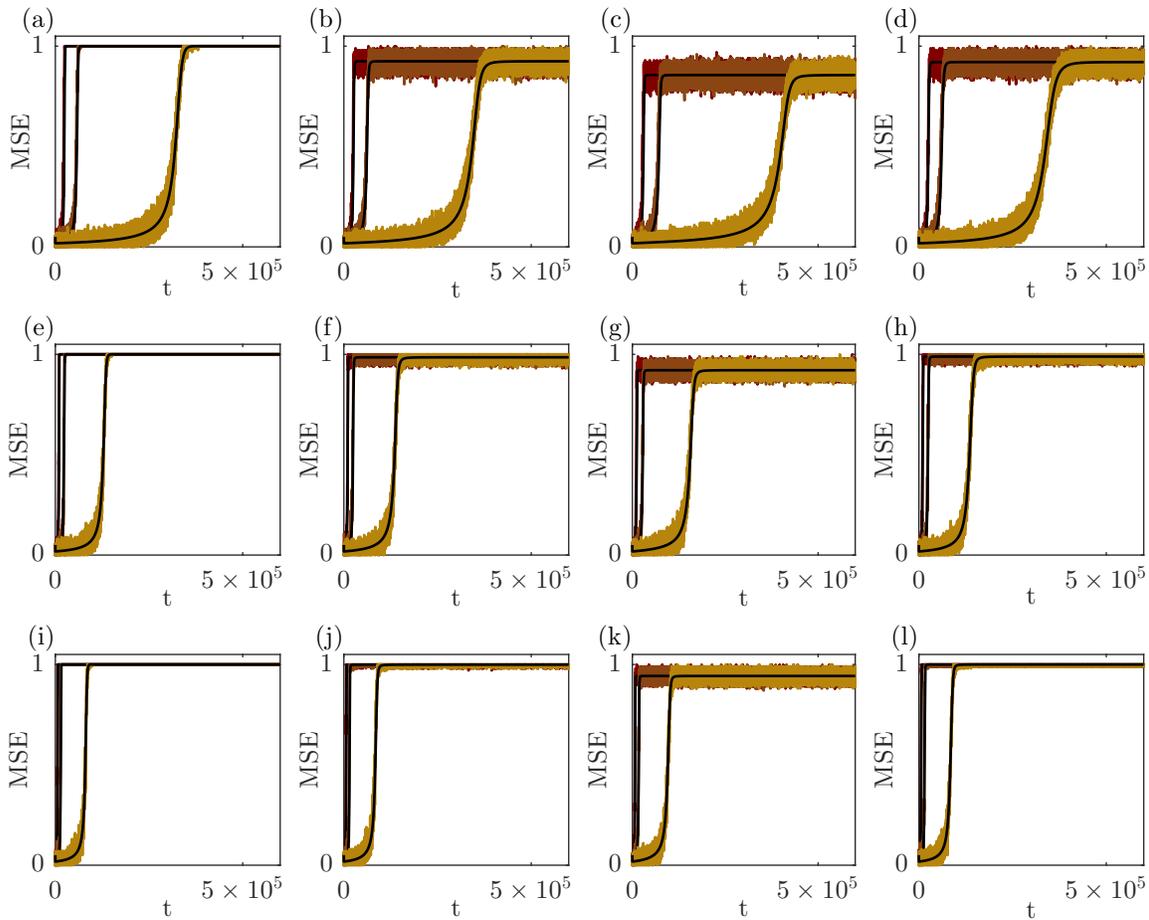}
\caption{Mean squared error (MSE) comparison of the stochastic and the deterministic models. For the stochastic model, the results are averaged across 100 network realizations.  \textbf{(a)} Regular graph. \textbf{(b)} ER graph. \textbf{(c)} BA graph. \textbf{(d)} WS graph. \textbf{(a-d)} $b/c = 1.8$. \textbf{(e-h)} Same as \textbf{(a-d)}, only $b/c = 1.5$. \textbf{(i-l)} Same as \textbf{(a-d)}, only $b/c = 1.2$. \textbf{(a-l)} A lighter shade indicates lower parameter values. All graphs have 100 nodes and average degree 8. In each run we set the initial values $Y_i(0) = -4$ and $p_i(0) = 0.05$ for all $i$. We assume synchronous update. 
\label{fig:MC_examples}}
\end{figure*}

\subsection{Deterministic approximation}
\label{sec:Deterministic_Model}

We approximate the stochastic model (\ref{eq:payoff}) by a deterministic model (under the same behavioral update), where the random variables are substituted with their respective expectations
\begin{align}
\mathrm{y}_i(t)&=b \sum_j \frac{A_{ij}}{d_i} \mathrm{p}_{j}(t) - c z_i  \mathrm{p}_{i}(t).
 \label{eq:deterministic_compact}
\end{align}
In (\ref{eq:deterministic_compact}), $z_i$ is defined as 
\begin{align*}
z_i=\sum_j A_{ji} / d_j.
\end{align*} 
This quantity acts as a local centrality measure of a node, with node $i$ being more ``important'' if it has many neighbors, and the neighbors themselves have few neighbors. In our model of interactions, this node would be called upon rather frequently. The measure, that we refer to as ``neighborhood importance index'', reflects the role of network topology in the promotion and stability of cooperation. When considering random walks on complex networks, one can show that $z_i$ is exactly the sum of the jump probabilities towards node $i$ from its neighbors \cite{chen2004local}.

When written in vector form, for  (\ref{eq:deterministic_compact}) we have 
\begin{align*}
\mathbf{y}(t) = \mathbf{\Theta}\cdot\mathbf{p}(t), 
\end{align*}
where $\Theta_{ii}= -c z_i$, and $\Theta_{ij}=b \frac{A_{ij}}{d_i}$, for $i\neq j$. 

In Fig.~\ref{fig:MC_examples} we present the comparison between the stochastic and the deterministic model in terms of the mean squared error (MSE) between the realizations of the individual cooperation probabilities and their deterministic counterparts (analytical solutions of the deterministic model). We observe that in the time limit the steady state behavior of both models is almost identical. In addition, the steady state solution depends only on the benefit-to cost ratio $b/c$, and the particular values of $b$ and $c$ only determine the rate of convergence, i.e the duration of the transient regime.

\section{Results}
\label{sec:Results}

Here, we address in more details the issues of cooperation in relation to the network topology. We thereby highlight the role of the neighborhood importance index $z$ (more precisely its distribution over the network nodes). In particular, we describe the steady state behavior of the deterministic model and derive important properties related to the existence and stability of cooperation.
\subsection{Steady state behavior} 
The update rule (\ref{eq:update}) yields the following set of iterative equations for $i=1,\ldots,N$
\begin{align*}
\mathrm{p}_i(t+1) &= \mathrm{f}\left(\mathbf{\mathrm{Y}}_i(t-1)+\mathbf{\Theta}_i\cdot\mathbf{p}(t)\right),
\end{align*}
where $\mathbf{\Theta}_i$ is the $i-$th row of $\mathbf{\Theta}$. 
In steady state it has to be fulfilled 
\begin{align*}
\mathrm{p}^*_i&=\mathrm{f}\left(\mathrm{f}^{-1}\left(\mathrm{p}^*_i\right)+\mathbf{\Theta}_i\mathbf{p}^*\right),
\end{align*}
for $i=1,\ldots,N$. By applying the inverse map we get 
\begin{align}
\mathrm{f}^{-1}\left(\mathrm{p}_i^*\right)&=\mathrm{f}^{-1}\left(\mathrm{p}_i^*\right)+\mathbf{\Theta}_i\mathbf{p}^*.
\label{eq:steady_state_individual_inverse}
\end{align}
The above requires $\mathrm{y}^*_i\doteq\mathbf{\Theta}_i\mathbf{p}^*=0$, unless either $\mathrm{p}_i^*=1$ (i.e. $\mathbf{\mathrm{Y}}_i^*=\mathrm{f}^{-1}\left(\mathrm{p}_i^*\right)=\infty$), or $\mathrm{p}_i^*=0$ (i.e.$\mathbf{\mathrm{Y}}_i^*=-\infty$). 

It is easy to verify that if there exists $i$ such that $\mathrm{p}^*_i=0$, then the same is true for all $i\in\mathcal{N}$. Indeed, when $p^*_i=0$, then from (\ref{eq:deterministic_compact}) it must hold that either: 1) $\mathrm{y}^*_i>0$, or: 2) $\mathrm{p}^*_j=0$ for all $j$ in the neighborhood of $i$, $j\in \mathcal{N}_i$. The condition 1 implies $\mathrm{p}^*_i=1$, which is a contradiction. The condition 2 yields $\mathrm{p}^*_i=0$ for all $i\in\mathcal{N}$ by repeating the same argument to the nodes in the neighborhood of $i$, until all nodes are reached. We note that this case is also covered by the requirement $\mathbf{\Theta}_i\mathbf{p}^*=\mathbf{0}$, with the solution $\mathbf{p}^*=\mathbf{0}$. 
Hence, a steady state solution fulfills $\mathbf{p}^*\in \mathbf{0}\cup (0\:\:1]^N$ and is thereby characterized by non-negative steady state payoffs $\mathrm{y}^*_i\geq 0$. 
We note that a steady state $\mathbf{p}^*=\mathbf{0}$ is also reached whenever the initial conditions are $\mathbf{p}(1)=\mathbf{0}$. We will, however, exclude this trivial possibility in the analysis that follows.

In steady state, the nodes may thus be attributed to two (disjoint) sets, $\mathcal{W} = \left\{ w \in \mathcal{N} : \mathrm{y}^*_w=0  \right\}$ and $\mathcal{S} = \left\{ s \in \mathcal{N} : \mathrm{y}^*_s>0  \right\}$, depending on the steady state payoff $\mathrm{y}^*_i$. As a consequence of (\ref{eq:steady_state_individual_inverse}), the nodes in $\mathcal{S}$ are further characterized by $\mathrm{p}^*_i=1$, while the nodes in $\mathcal{W}$ may take both values $\mathrm{p}^*_i=1$ and $\mathrm{p}^*_i<1$, depending on the network parameters. We will refer to the nodes in the sets $\mathcal{W}$ and $\mathcal{S}$ as ``weak'', respectively ``strong'' nodes, with an intention to emphasize their role in the bifurcation analysis performed later. Accordingly, there are two sets of relations that have to be satisfied 
\begin{align}
0&= \frac{b}{d_i} \sum_j A_{ij} \mathrm{p}_j^* - c z_i\mathrm{p}^*_i, \:\: i\in \mathcal{W}\nonumber\\
\mathrm{y}^*_i &= \frac{b}{d_i} \sum_j A_{ij} \mathrm{p}_j^* - c z_i, \:\: i\in \mathcal{S}.
\label{eq:optimization_model_simplified}
\end{align}  
Note that in (\ref{eq:optimization_model_simplified}) the sets $\mathcal{W}, \mathcal{S}$, the steady state values $\mathrm{p}_i^*,\:i\in\mathcal{W}$ and the constants $\mathrm{y}^*_i, \:i\in \mathcal{S}$ are unknown. 

\subsection{Properties of the model} 

In the following we derive some important properties of the model. In particular, we investigate the necessary and sufficient conditions for the spread and stability of cooperation, in relation to the network topology. \smallskip

\begin{figure}[t!]
\includegraphics[width=8.6cm]{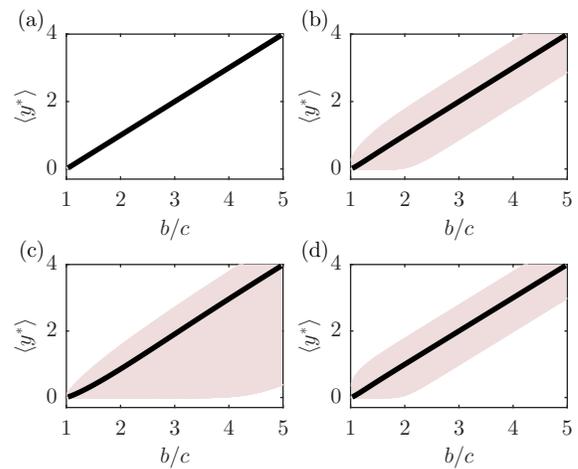}
\caption{Average steady state payoff $\langle\mathrm{y}^*\rangle$ as a function of $b/c$ (solid line), averaged over 100 graph realizations. The regions (in lighter shade) enclose the area between the minimum and the maximum node payoff observed in the network. \textbf{(a)} Regular graph. \textbf{(b)} ER graph. \textbf{(c)} BA graph. \textbf{(d)} WS graph. \textbf{(a)-(d)} All graphs have $1000$ nodes and average degree $8$.}
\label{fig:ci_payoff}
\end{figure}

\noindent\textit{\textbf{1}.~Robustness to exploitation:} The non-negativity of the individual steady state payoffs $\mathrm{y}_i^*$ in  (\ref{eq:optimization_model_simplified}) has an important implication on the promotion of cooperation in networks as it ultimately protects the individual nodes from exploitation by the rest of the network. This is, in general, at contrast to other mechanisms based on general reciprocity where the anonymity of donors and receivers makes it difficult to single out and punish defectors, leaving the nodes vulnerable to exploitation. Fig.~\ref{fig:ci_payoff} illustrates the range of the individual steady state payoffs and the average network payoff $\langle y^* \rangle$ as a function of the benefit/cost ratio $b/c$, for the different network models. 

\noindent\textit{\textbf{2.}~Necessary condition for the existence of cooperation:} It is easy to show that $b/c < 1$ implies $\mathrm{p}_i = 0$ for all $i\in\mathcal{N}$. Indeed, if there exists $i$ such that $\mathrm{p}^*_i>0$, then the total steady state network payoff
\begin{align} \label{eq:network-payoff}
\sum_i \mathrm{y}^*_i&=b \sum_i \sum_j \frac{A_{ij}}{d_i} \mathrm{p}^*_j - c \sum_i \sum_j \frac{A_{ji}}{d_j}  \mathrm{p}^*_i \nonumber \\
&= \left( b - c \right) \sum_i  z_i \mathrm{p}^*_{i}, 
\end{align}
is strictly negative, implying that there is some $i$ for which $\mathrm{y}^*_i < 0$ (contradiction). Hence, $b/c\geq 1$ is a necessary condition for the existence of cooperators (nodes with $\mathrm{p}^*_i>0$). 

\noindent\textit{\textbf{3.}~Promotion of cooperation:} We observe that when $b/c>1$, the steady state probabilities are strictly greater than $0$, $\mathrm{p}_i^*>0$ for all $i\in \mathcal{N}$. Indeed, if there exists $i$ such that $p^*_i=0$ then, as already discussed, it must hold that $\mathrm{p}^*_i=0$, for all $i\in\mathcal{N}$. This, however, would yield a total network payoff $\sum_{i}\mathrm{y}^*_i=0$, which contradicts (\ref{eq:network-payoff}).  

\noindent\textit{\textbf{4.}~Sufficient condition for the existence of unconditional cooperators (strong nodes):} When $b/c>1$, there is always at least one strong node in the network. This follows directly from the observation that when $b/c>1$ the RHS of (\ref{eq:network-payoff}) is strictly greater than zero, which implies that there is at least one $i$ for which $\mathrm{y}^*_i > 0$ and $\mathrm{p}^*_i = 1$. Combined together, \textit{property 3.} and \textit{property 4.} state that, as a consequence of the update rule \ref{eq:update}, the network nodes cooperate with the maximum possible probability, such as their pay-off is non-negative. In other words, they are not exploited by their environment. 

\noindent\textit{\textbf{5.}~Necessary condition for the existence of unconditional cooperators (strong nodes):} The condition $z_i > b/c$ implies $\mathrm{p}_i^*<1$, which follows by substituting $\sum_j \frac{A_{ij}}{d_i} \mathrm{p}^*_j\leq 1$ in (\ref{eq:optimization_model_simplified}). In other words, a necessary condition for $i$ to be strong (i.e. unconditional cooperator $\mathrm{p}_i^*=1$) is $z_i\leq b/c$.  

\noindent\textit{\textbf{6.}~Necessary and sufficient condition for full network cooperation:} We show that 
\begin{align}
b/c \geq z_{\max},
\label{eq:full_coop_threshold}
\end{align}
 where $z_{\max}$ is the largest neighborhood importance index in the graph, $z_{\max}=\max_i (z_i)$, is both necessary and sufficient for all nodes to be strong (full network cooperation). 

We note that the proof that $\mathrm{p}_i^*=1$, $\forall i\in \mathcal{N}$, implies $b/c \geq z_{\max}$, follows directly from \textit{property 3}. To prove the converse, we use contradiction. We first define $\mathrm{p}_{\min}=\inf \mathrm{p}_i^*, \: i\in \mathcal{N}$, and set $b/c$ to be greater than one ($b/c>1$ being the prerequisite for cooperative behavior). 

Now, let us assume that the converse is not true, that is $ b/c \geq z_i $ for all $i$, and there exists some $i$ such that $\mathrm{p}^*_i < 1$. Under this assumption, for all $i$ we would have:
\begin{align*}
\mathrm{y}^*_i &= b \sum_j \frac{A_{ij}}{d_i} \mathrm{p}^*_j -c z_i  \mathrm{p}^*_i, \\
& \geq b \sum_j \frac{A_{ij}}{d_i} \mathrm{p}^*_j -c \frac{b}{c}  \mathrm{p}^*_i,
\end{align*}
which implies
\begin{align*}
\mathrm{p}^*_i  + \frac{\mathrm{y}^*_i}{b} \geq  \sum_j \frac{A_{ij}}{d_i} \mathrm{p}^*_{j}.
\end{align*}
However, for those $i$ satisfying  $\mathrm{p}^*_i<1$, we know that $\mathrm{y}^*_i = 0$, implying
\begin{align}
& \mathrm{p}^*_i \geq \sum_j \frac{A_{ij}}{d_i} \mathrm{p}^*_j \geq \mathrm{p}_{\min} \label{eq:pmin}.
\end{align}
For all $i$ satisfying $b/c > z_i$,  (\ref{eq:pmin}) holds with strict inequality, whereas those $i'$ for which $b/c = z_{i'}$ must satisfy $\mathrm{p}^*_{i'} = \mathrm{p}_{\min}$. This, however, can hold if and only if  the nodes corresponding to these indices are only linked to each other, i.e. form a connected component. In that case $z_{\max} = 1 = b/c$ which contradicts the assumption $b/c > 1$. Hence, the converse must also be true, which concludes the proof.

\subsection{Derivation of the steady state solution}

\begin{figure}[t!]
\includegraphics[width=8.6cm]{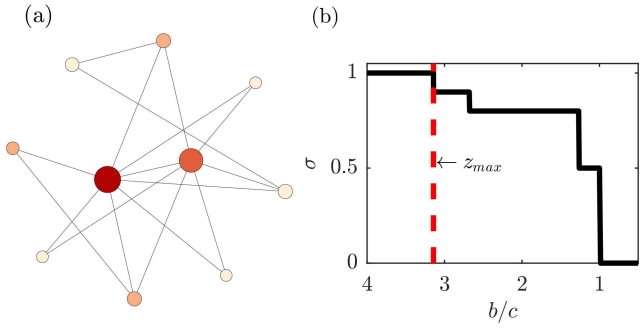}
\caption{ \textbf{(a)} A visualization of a small random graph with 10 nodes. The nodes are colored according to the order of switching from strong to weak when decreasing the $b/c$ ratio -- a darker color indicates higher switching propensity. The node size is proportional to the index $z_i$. \textbf{(b)} Fraction of strong nodes $\sigma$ as a function of the (decreasing) $b/c$ ratio. \label{fig:bifurcation}}
\end{figure}

Having addressed the general steady state behavior of the model and derived its most important properties, here we provide bifurcation analysis to determine the steady state solution of (\ref{eq:optimization_model_simplified}). To do so, we keep the cost $c$ fixed and vary the benefit $b$ in order to determine their influence on the system. Fig. \ref{fig:bifurcation} serves as a graphical illustration for the analysis.

We start with the remark that $b/c\geq z_{\max}$ ensures that all nodes are unconditional cooperators (i.e. strong), $\mathrm{p}_i^* = 1$, for all $i\in\mathcal{N}$. This follows directly from \textit{necessary and sufficient condition for full network cooperation (property \textbf{6}.)}. We denote $b_1 = c z_{\max}$ and introduce $\mathcal{W}_1$ as the set of nodes with indices $ w_1 \in \mathcal{W}_1$ satisfying $z_{w_1}=b_1/c$ (note that $\mathcal{W}_1$ may have more than one element). For $i\in \mathcal{W}_1$ the payoff becomes $y_i^*=0$, while it is still $\mathrm{p}_i^*=1$. By reducing $b$ beyond $b_1$, for $i\in \mathcal{W}_1$ the probability of cooperation becomes $\mathrm{p}^*_i<1$ (i.e the nodes become weak), as $\mathrm{p}^*_i=1$ would imply a payoff $\mathrm{y}_i^*<0$ (contradiction). The value(s) $\mathrm{p}^*_i$ are determined from (\ref{eq:optimization_model_simplified}), by plugging in $\mathrm{p}^*_i<1$ for $i\in \mathcal{W}_1$. Thus, $b_1$ is the first bifurcation point and the nodes in the set $\mathcal{W}_1$ are the first to become weak, i.e. to break up with unconditional cooperation. The remaining nodes $\mathcal{S}^{(1)}=\mathcal{N} \setminus \mathcal{W}_1$ are still strong.  

We proceed by using induction to determine the set of weak nodes $\mathcal{W}^{(n+1)}=\bigcup_{i=1}^n\mathcal{W}_i$ and the (remaining) strong nodes $\mathcal{S}^{(n+1)}=\mathcal{N} \setminus \mathcal{W}^{(n+1)}$ for any $b$ in the interval between two bifurcation points, $b_n>b\geq b_{n+1}$. After substituting  $\mathrm{p}^*_i=1$ for the nodes in  $\mathcal{S}^{(n+1)}$, we determine $\mathrm{p}^*_i$ for the nodes in $\mathcal{W}^{(n+1)}$ from the remaining equations. 

By applying some simple algebra to reorganize the equations for the nodes in the set $\mathcal{W}^{(n)}$, for all $b$ in the interval between $b_n$ and the next bifurcation point $b_{n+1}$, $b_n>b\geq b_{n+1}$, we obtain 
\begin{align*}
\sum_{j \in \mathcal{S}^{(n)}} \frac{b}{c z_i d_i}A_{ij} &= p_i^* - \sum_{j \in \mathcal{W}^{(n)}} \frac{b }{c z_i d_i} A_{ij} p_j^*,\:\:\:\:\:i\in \mathcal{W}^{(n)}.
\end{align*}
The last equation can be written in matrix form 
\begin{align}
\mathbf{a} = \left(\mathbf{I} - \mathbf{A}_{\mathcal{W}^{(n)}} \right) \mathbf{p}^*,
\label{eq:matrix_bifurcation}
\end{align}
where the vector $\mathbf{a}$ is the vector of sums appearing on the left-hand side of (\ref{eq:matrix_bifurcation}), while $\mathbf{A}_{\mathcal{W}^{(n)}}$ is a weighted version of the neighborhood matrix of the subgraph associated with $\mathcal{W}^{(n)}$. The solution is unique provided that the matrix $\mathbf{I} - \mathbf{A}_{\mathcal{W}^{(n)}}$ is nonsingular, i.e. the inverse $\left(\mathbf{I} - \mathbf{A}_{\mathcal{W}^{(n)}} \right) ^ {-1}$ exists. In that case the solution reads 
\begin{align*}
\mathbf{p}^* = \left(\mathbf{I} - \mathbf{A}_{\mathcal{W}^{(n)}} \right) ^ {-1} \mathbf{a}.
\end{align*}
We note that in reality, and particularly in large networks, the nonsingularity condition for $\mathbf{I} - \mathbf{A}_{\mathcal{W}^{(n)}}$ is fulfilled almost certainly.

\subsection{Alternate projection method for the steady state solution}

The bifurcation analysis provides an analytical solution to the steady state cooperation probabilities for each $b/c>1$. By the definition of steady state, the same solution would be obtained by starting from an arbitrary initial condition $\mathbf{p}(0)\in (0\:\:1]^N$ and letting the network evolve according to our payoff and update rules. Note that the steady state conditions may be reformulated as 
\begin{align*}
\mathrm{y}^*_i &= b \sum_j \frac{A_{ij}}{d_i} \mathrm{p}^*_j -c z_i  \mathrm{p}^*_i \nonumber \\ 
0 &= \mathrm{y}^*_i \left( 1 - \mathrm{p}^*_i \right), \label{eq-nonlin}
\end{align*} 
resulting in a nonlinear (quadratic) system of $2N$ equations with $2N$ variables in total. Hence, a simplified, iterative approach based on the alternate projection method can be used for finding the steady state solution. This method may be summarized as follows:
\begin{enumerate}
\item Set $\mathrm{y}^*_i = 0$ for all $i$ satisfying the condition $ z_i \geq b/c$. Set $\mathrm{p}^*_i = 1$ for the remaining nodes. Solve the $N$-dimensional linear system to find the remaining $\mathrm{p}_i^*$ and $\mathrm{y}^*_i$ ($N$ unknowns in total)
\item For all $i$ satisfying $\mathrm{y}^*_i < 0$ in the obtained solution, set $\mathrm{y}^*_i = 0$ and let their corresponding $\mathrm{p}^*_i$ to be unknown. Solve again the corresponding linear system of $N$ equations with $N$ unknowns in total.
\item Repeat steps $1.$ and $2.$ until there are no $\mathrm{y}^*_i < 0$.
\end{enumerate}

\subsection{Implications of the model}

\begin{figure*}[t!]
\includegraphics[width=17cm]{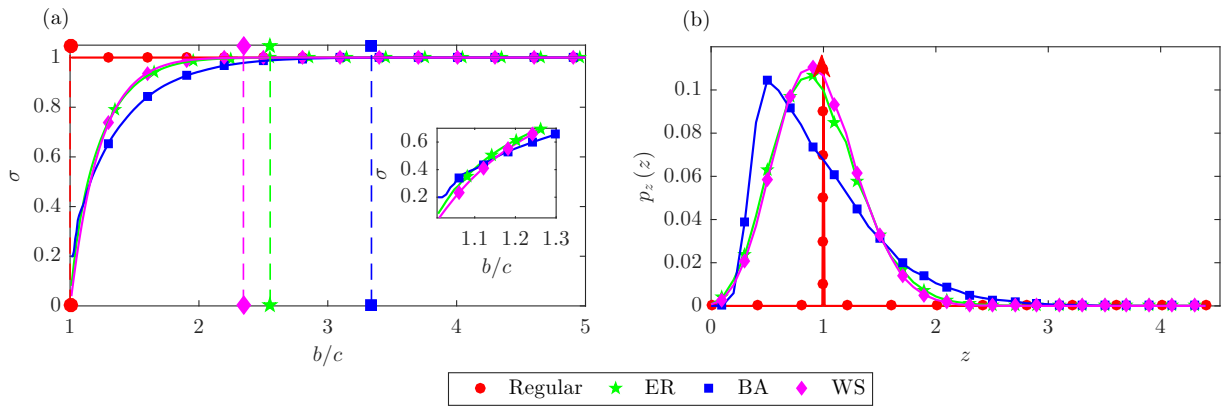}
\caption{\textbf{(a)} Fraction of unconditional cooperators $\sigma$ as a function of $b/c$ for Regular, ER, BA, and WS graphs. Dashed lines indicate the threshold for full network cooperation according to equation (\ref{eq:full_coop_threshold}). 
\textbf{(b)} Probability density function (pdf) of the index $z$ for the same random graphs.
We note that for the Regular graph the pdf is concentrated on a single point (i.e the index $z$ is the same across all nodes, $z=1$).
\textbf{(a-b)} The results are obtained by averaging over 100 different graph realizations. All graphs have $1000$ nodes and average degree $8$.  }
\label{fig:density+fraction_coop}
\end{figure*}

While the connection is not imminent, the addressed model may still be framed in the context of evolutionary game theory, by associating the accumulated payoff with fitness of the individual nodes. In this sense, our main contribution can be related to the results for the known rules for evolution of cooperation (see Table \ref{table:coop-mechanism}). In particular, it can be seen as a rule for cooperation based on generalized reciprocity on graphs. One has to be careful, however, when comparing the different rules, as they usually arise from different setups and are dependent on the interaction model/game update. For example, the model in \cite{Barta-2011} assumes pairwise node interactions where each node pair is chosen on random during each iteration, yielding a uniform distribution on the number of interaction instances per node. In our model, on the other hand, some nodes (i.e. those with a large $z$-index) engage more regularly in the interactions with their neighbors, thus reflecting the role of network structure. With this in mind, and after accounting for the differences in the interaction model, a natural connection may be made between the presented mechanism based on generalized reciprocity on graphs, and the mechanisms based on indirect reciprocity \cite{Nowak-1998}, network reciprocity (revisited) \cite{Konno-2011}, and generalized reciprocity \cite{Barta-2011}, as highlighted in Table \ref{table:coop-mechanism}.

We turn the attention to Fig.~\ref{fig:density+fraction_coop}a which sheds light on the effect of network structure on the promotion of cooperation under the addressed model. As an indicator for the level of network cooperation we consider the steady state fraction of unconditional cooperators $\sigma$ as a function of the benefit/cost ratio $b/c$ ($b>c$ is the prerequisite for cooperation). The figure reveals that the Barabasi-Albert (BA) scale-free graph requires the largest $b/c$  for full network cooperation to take place (all nodes unconditional cooperators), followed by the Erdos-Renyi (ER) graph and the Watts-Strogatz (WS) small-world graph. In contrast, for a small $b/c$ ratio, the BA graph has the largest fraction of cooperators among these three graph types. The Regular graph presents itself as the most supportive to cooperation, as $b>c$ implies full, unconditional cooperation on network level.

The reason for this behavior may be directly inferred from Fig.~\ref{fig:density+fraction_coop}b, where we depict the probability density function (pdf) of the index $z$ across the network nodes for the same random graphs. We recall that that the individual cooperative behavior in the network is determined by this index. The nodes with a higher value of it are in ``less favorable'' position in the network, as they will be called upon more often in our interactions model. For higher values of the benefit-to-cost ratio $b/c$, the global cooperative behavior is dominated by the tail (the right-hand side) of the distribution, i.e. the fraction of network nodes with high values of the $z$-index. This explains exactly the lower fraction of unconditional cooperators in the BA graph in this regime, as compared to the other random graph configurations. On the other hand, the behavior in the low $b/c$ regime is determined by the left-hand side of the distribution, where again a large mass is distributed in the case of the BA graph. As a result, in this regime we find higher fraction of unconditional cooperators in the BA graph compared to the ER and the WS graph, as shown in the box of Fig.~\ref{fig:density+fraction_coop}a.    

\begin{table*}[t]
\caption{Cooperation Mechanisms}
\vspace{0.5cm}
\label{table:coop-mechanism}
\begin{tabular}{l|l|l}
\textbf{Mechanism} & \textbf{Rule} & \textbf{Note} \\
\hline
 Kin selection & $b/c > 1/g$ & $g$ denotes the probability that two agents share a gene \cite{Hamilton-1964}.   \\
 & &  \\
\hline
 Group selection & $b/c > 1+ n/m$    & $n$ and $m$, are respectively, the maximum group size       \\
 & &  and number of groups \cite{Traulsen-2006}. \\
\hline
 Direct reciprocity & $b/c > 1/w$ & $w$ is the probability that a game will last one more round \cite{Trivers-1971}.       \\
& &  \\
\hline
 Network reciprocity & $b/c > \langle d \rangle$ & $\langle d \rangle$ is the average number of neighbors \cite{Ohtsuki-2006}. \\
  & &  \\
\hline\hline
\rowcolor{Gray} Network reciprocity  & $b/c > \langle d_2 \rangle$ & $\langle d_2 \rangle$ is the average degree of the nearest neighbors \cite{Konno-2011}. \\
\rowcolor{Gray} (revisited) & & \\
\hline
 \rowcolor{Gray} Indirect reciprocity & $b/c > 1/q$ & $q$ is the probability to know someones reputation \cite{Nowak-1998}. \\
\rowcolor{Gray}  & & \\
\hline
\rowcolor{Gray} Generalized reciprocity & $b/c > (v+n^2)/(v-n^2)$    & $v$ denotes the number of interactions, and $n$ is the group size \cite{Barta-2011}. \\
\rowcolor{Gray} & &  \\
\hline
\rowcolor{Gray} Generalized reciprocity & $b/c \geq z_{\max}$ & $z_{\max}$ is the maximum neighborhood importance index.\\
\rowcolor{Gray} on graphs & &  \\
 \hline
\end{tabular}
\end{table*}

\subsection{Stability of cooperation} 

So far, we addressed the case when all nodes were subject to the same behavioral mechanism (\ref{eq:update}) to update their individual probabilities of cooperation. In the following, we assume that a fraction $\delta = D/N$ of the nodes become unconditional defectors ($\mathrm{p}_i(t)=0$), independent on their accumulated payoff. The aim is to assess the robustness of the (in general cooperative) behavioral update to the presence of defectors. 

The bifurcation analysis as well as the alternate projection method can be easily accommodated to account for this characteristic. Additionally, several properties derived for the original case can be generalized to this setup. Particularly, in the presence of unconditional defectors, the \textit{necessary condition for existence of unconditional cooperators} (\textit{property \textbf{5}.}), becomes $ \frac{d_i}{q_i} z_i \leq b/c$, where $q_i$ is the number of neighbors of $i$ that are not unconditional defectors. In addition, the \textit{necessary and sufficient condition for full network cooperation} (\textit{property \textbf{6}.}) now reads $b/c \geq \max_i \frac{d_i}{q_i} z_i $. The remaining properties can not be easily generalized as they depend on the initial selection of defecting nodes. 

\begin{figure}[t!] 
\includegraphics[width=8.6cm]{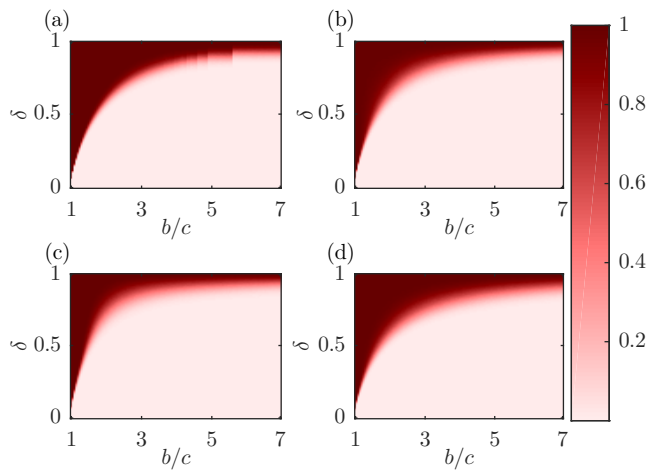}
\caption{Heat map for the fraction of defective instances as a function of $\delta = D/N$ and $b/c$. $D$ out of $N$ nodes are randomly chosen to be unconditional defectors and the results are averaged over 100 realizations. \textbf{(a)} Regular graph. \textbf{(b)} ER graph. \textbf{(c)} BA graph. \textbf{(d)} WS graph. \textbf{(a)-(d)} All graphs have $100$ nodes and average degree $8$. Results are averaged across 100 graph realizations. \label{fig:heatmap_defectors} }
\end{figure}

Numerical results are summarized in Fig.~\ref{fig:heatmap_defectors} where we plot a heat map for the fraction of ``defective instances'' as a function of $\delta$ and $b/c$. By defective instance we understand the absence of a strong node in the network in steady state. Interestingly, the BA graph presents itself as the most robust to an invasion of defectors, as it registers the smallest number of defective instances $\delta$ for the same ratio $b/c$. The ER graph provides to be slightly less robust than the BA graph. As least robust to an invasion of defectors come the WS and the Regular graph, which present similar behavior in this setup. The reason for this behavior becomes apparent if we,  again, look at the distribution of the index $z$ in different random graphs. Since in this case we are interested in the number of defective instances (i.e. graph realizations without unconditional cooperators in the network), decisive is the shape of the left-hand-side of the pdf of $z$. As the existence of unconditional defectors in the neighborhood of a node $i$ decreases the payoff of node $i$, its influence on the cooperative behavior of node $i$ may be understood as an ``effective'' increase of the value of the index $z_i$. This, on the other hand may drive unconditional cooperators to extinction (a defective distance is declared for the particular configuration). As depicted in Fig.~\ref{fig:density+fraction_coop}b, due to the higher fraction of nodes with small values of the index $z$, the BA graph is on average less affected by this condition, i.e. it is more robust to an invasion of defectors.

\subsection{Accounting for restricted memory}
The state-based behavioral rule (\ref{eq:update}) implicitly assumes that each past event is valued equally. However, sometimes the agents (nodes) have short memory, and thus may give higher importance to recent interactions. We can account for this effect by adding a weight $r$, $0 \leq r \leq 1$, to the accumulated payoff, i.e.
\begin{align*} 
Y_i(t) &= r Y_i(t-1) + \mathbf{\Theta}_i\mathbf{p}(t). 
\end{align*}
We note that the case $r=1$ corresponds to the update (\ref{eq:update}). When $r<1$, the steady state probability for cooperation for node $i$ is
\begin{equation} \label{eq:steady-state-memory}
 \mathrm{p}^*_i = \mathrm{f} \left( \frac{\mathbf{\Theta}_i\mathbf{p}^*}{1-r}   \right) = \mathrm{f} \left( \frac{\mathrm{y^*_i}}{1-r}   \right). 
\end{equation}
 In this case, the accumulated payoff converges to a finite value since the steady state payoffs represent converging geometric series, implying $0 < \mathrm{p}^*_i < 1$. We point out that, with the introduction of the weight $r$, exploitation can not be prevented by employing the update rule (\ref{eq:update}) for small enough $b/c$ ratios, as in some cases it may happen that $\mathrm{y}^*_i < 0$ for some $i$. This can be seen by applying the inverse map to (\ref{eq:steady-state-memory}), as we get
\begin{align*} 
\mathrm{y}^*_i &= \left( 1-r \right) \mathrm{f}^{-1} \left(\mathrm{p}^*_i \right),
\end{align*}
which can be negative if $\mathrm{f}^{-1} \left(\mathrm{p}^*_i \right)$ is negative. This happens, for example, if $\mathrm{f}\left( \cdot \right)$ is the standard logistic function, with steepness $k = 1$ and midpoint $\omega_0 = 0$, and $\mathrm{p}^*_i < 1/2$.

However, there is a threshold $\beta^*$ for which if $b/c \geq \beta^*$ exploitation is prevented. As can be seen from equation (\ref{eq:steady-state-memory}) the threshold depends on the underlying network type and the parameter $r$, and can not be easily derived. Nevertheless, we can provide its lower and upper bounds, $\beta_{LB}$ and $\beta_{UB}$ that are independent from $r$. In particular, we can safely assume that $\beta_{LB} = 1$ because, as previously said, $b/c > 1$ is a prerequisite for cooperation. To derive the upper bound $\beta_{UB}$, we again assume the existence of the minimal probability for cooperation $\mathrm{p}_{\min}$ (its presence is a valid assumption since, from (\ref{eq:steady-state-memory}), we know that $\mathrm{p}_{i} > 0$ for all $i$). Then, from (\ref{eq:optimization_model_simplified}), it follows that $\mathrm{y}_i \geq b \mathrm{p}_{\min} - c z_i \mathrm{p}_i$. Therefore, in order for $y_i$ to be nonnegative for all $i$ it must be that
\begin{align*}
b \mathrm{p}_{\min} \geq c z_i \mathrm{p}_{i} \geq c z_i \mathrm{p}_{\min}
\end{align*}
for all $i$, or $b/c \geq z_{\max}$. Interestingly, this is the same value as the condition for full network cooperation in the infinite memory case. When the network is regular, the bounds coincide, thus leading to $\beta^* = 1$, a result which is independent of $r$.

Finally, it is worth mentioning that, when $r = 0$, $ \mathrm{f}(\cdot)$ is very steep (such that $\mathrm{p}_{i}(t+1) = 1$ if $\mathrm{y}_{i}(t) > 0$, and $\mathrm{p}_{i}(t+1) = 0$, otherwise)  and setting $N = 2$, we get the tit-for-tat strategy in the iterated prisoner's dilemma.

\section{Discussion} 
We argue that the (stochastic) equations (\ref{eq:payoff}) and (\ref{eq:update}) provide a realistic model for the dynamics of network interactions in a large plethora of real-life networks. Under this model, network cooperation comes as a result of the inherent feature of the behavioral mechanism that prevents participating nodes from being exploited by the environment. In particular, the simple update rule (\ref{eq:update}) requires minimal cognitive abilities and very little information retention and retrieval as the decisions of individuals to cooperate or not, only depend on their internal state which captures the past experience of their interactions with the network. The internal state may mirror fitness in biological systems, wealth or well-being in animal and human societies, or battery level (energy) in artificial systems (e.g. wireless ad-hock networks). Due to its simplicity, this behavioral mechanism is more likely to evolve in real networks than, e.g. direct or indirect types of reciprocity, which require much more specific memory, cognitive ability and effort.   

A corollary of our findings is that, under the addressed model for network interactions (\ref{eq:payoff}), and the behavioral mechanism (\ref{eq:update}), the BA scale-free graph is the most inhibitive to cooperation, followed by the ER graph and the WS small-world graph, while the regular graph presents itself as the most supportive for cooperation (as displayed in Fig.~\ref{fig:density+fraction_coop} and Fig.~\ref{fig:ci_payoff}). The picture is inverted in the presence of pure defectors in the network, as then, the BA graph provides the highest degree of robustness, followed by the ER graph (as captured by Fig.~\ref{fig:heatmap_defectors}). 

We point out that, while most of our conclusions are along the same lines in \cite{Konno-2011, Nowak-1993, Fu-2009}, they do not contradict the findings in \cite{Santos-2005, Santos-2006_2} which suggest that, under the model addressed there, scale-free networks enhance cooperation. We argue that the apparent inconsistencies found in the literature, may be attributed to the differences in the way that network interactions are modeled. Therefore, it is important that all  findings in this context are, in general, interpreted in light of the specifics of the addressed model only.

\vspace{2pt}
\section*{Acknowledgement}

This research was supported in part by DFG through grant ``Random search processes, L\'evy flights, and random walks on complex networks''.

\bibliography{coop-bib}

\end{document}